\documentclass[useAMS,usenatbib,usegraphicx]{mn2e}

\usepackage{amssymb} 
\usepackage{textcomp} 
\usepackage{appendix}
\usepackage{amsmath} 
\usepackage{fixltx2e} 
\usepackage{multirow} 
\usepackage{enumerate} 
\usepackage{mathptmx} 
\usepackage{lscape}  
\usepackage[colorlinks=true,citecolor=blue,linkcolor=blue,urlcolor=blue]{hyperref} 
\usepackage{units} 

\newcommand{\kms}{\,\rm km\ s^{-1}}
\newcommand{\ergs}{\,\rm erg\ s^{-1}}

\newcommand{\mic}{\,\mbox{$\mu$m}}

\newcommand{\kev}{\,\rm keV}

\let\AAold\AA
\renewcommand{\AA}{\text{\AAold}}

\newcommand{\cm}{\,{\rm cm}}
\newcommand{\pc}{\,{\rm pc}}

\newcommand{\yr}{\,{\rm yr}}

\newcommand{\ryd}{\,{\rm Ryd}}

\newcommand{\K}{\,{\rm K}}

\newcommand{\zsun}{\,{\rm Z_{\odot}}}

\renewcommand{\mp}{m_{\rm p}}
\newcommand{\kB}{k}
\newcommand{\sth}{\sigma_{\rm Th}}

\newcommand{\hnu}{\langle h\nu\rangle}

\newcommand{\aion}{\alpha_{\rm ion}}

\newcommand{\ebv}{E_{\rm B-V}}

\newcommand{\nH}{n_{\rm H}}

\newcommand{\hi}{\text{H~{\sc i}}}
\newcommand{\hii}{\text{H~{\sc ii}}}

\newcommand{\aap}{A\&A}
\newcommand{\araa}{ARA\&A}
\newcommand{\apjl}{ApJ}
\newcommand{\apjs}{ApJS}
\newcommand{\apj}{ApJ}

\newcommand{\mnras}{MNRAS}
\newcommand{\rmxaa}{RevMexA\&A}

\newcommand{\ssr}{Space Science Reviews}
\newcommand{\physrep}{Physics Reports}
\newcommand{\aapr}{The Astronomy and Astrophysics Review}
\newcommand{\solphys}{Solar Physics}

\renewcommand{\v}{\text{v}}

\newcommand{\lion}{L_{\rm ion}}

\newcommand{\Nh}{N}
\newcommand{\Ntot}{N_{\rm tot}}

\newcommand{\prad}{P_{\rm rad}}
\newcommand{\pmag}{P_{\rm mag}}
\newcommand{\pgas}{P_{\rm gas}}
\newcommand{\pgasz}{P_{\rm gas,0}}

\newcommand{\frad}{F_{\rm rad}}

\newcommand{\sigbar}{{\bar \sigma}}
\newcommand{\taubar}{{\bar \tau}}
\renewcommand{\d}{{\rm d}}
\newcommand{\tc}{T_{\rm C}}
\newcommand{\pline}{P_{\rm line}}
\newcommand{\vturb}{\Delta v}

\newcommand{\cloudy}{{\sc cloudy}}

\newcommand{\xspec}{{\sc xspec}}
\newcommand{\xstar}{{\sc xstar}}
\newcommand{\titan}{{\sc titan}}

\title[Radiation Pressure Confinement]{Radiation Pressure Confinement -- III. The origin of the broad ionization distribution in AGN outflows}
\author[Jonathan Stern, Ehud Behar, Ari Laor, Alexei Baskin and Tomer Holczer]
{
Jonathan Stern$^{1,2}$\thanks{E-mail: stern@mpia.de}, 
Ehud Behar$^2$, Ari Laor$^2$, Alexei Baskin$^2$ and Tomer Holczer$^3$ \\
$^1$ Max-Planck-Institut f\"{u}r Astronomie, K\"{o}nigstuhl 17, D-69117 Heidelberg, Germany\\
$^2$ Physics Department, Technion -- Israel Institute of Technology, Haifa~32000, Israel\\
$^3$ School of Physics and Astronomy, Raymond and Beverly Sackler Faculty of Exact Sciences, Tel Aviv University, Tel Aviv~69978, Israel}
\begin{document}
\maketitle

\begin{abstract}
The winds of ionized gas driven by Active Galactic Nuclei (AGN) can be studied through absorption lines in their X-ray spectra. 
A recurring feature of these outflows is their broad ionization distribution, including essentially all ionization levels (e.g., Fe$^{0+}$ to Fe$^{25+}$). 
This characteristic feature can be quantified with the absorption measure distribution (AMD), defined as the distribution of column density with ionization parameter $|{\rm d}\Nh/\d\log\xi|$. 
Observed AMDs extend over $0.1\lesssim\xi\lesssim10^4$ (cgs), and are remarkably similar in different objects. 
Power-law fits $(|\d\Nh/\d\log\xi|\approx N_1\xi^a)$ yield $N_1=3\times10^{21}\cm^{-2}\pm0.4{\,\rm dex}$ and $a=0-0.4$. 
What is the source of this broad ionization distribution, and what sets the small range of observed $N_1$ and $a$? 
A common interpretation is a multiphase outflow, with a wide range of gas densities in a uniform gas pressure medium. 
However, the incident radiation pressure leads to a gas pressure gradient in the photoionized gas,
 and therefore to a broad range of ionization states within a single slab.
We show that this compression of the gas by the radiation pressure leads to an 
AMD with $|\d N/\d\log\xi|=8\times10^{21}\xi^{0.03}\cm^{-2}$, remarkably similar to that observed. 
The calculated values of $N_1$ and $a$ depend weakly on the gas metallicity, the ionizing spectral slope, the distance from the nucleus, the ambient density, and the total absorber column. Thus, radiation pressure compression (RPC) of the photoionized gas provides a natural explanation for the observed AMD. 
RPC predicts that the gas pressure increases with decreasing ionization, which can be used to test the validity of RPC in ionized AGN outflows.
\end{abstract}


\section{Introduction}
It is well established that Active Galactic Nuclei (AGN) drive winds of ionized gas (\citealt{Reynolds97, Crenshaw+99}). 
The outflows of the low luminosity AGN, the Seyfert galaxies, are readily observed through their rich X-ray and UV spectra (e.g. \citealt{Maran+96,Kaspi+01}), in which blue-shifted absorption lines allow for elaborate plasma diagnostics of the photo-ionized wind (for a review, see \citealt{Crenshaw+03}).
Seyfert winds are typically flowing at low to moderate velocities of $v \lesssim 1000$ km\,s$^{-1}$.

Although sensitivity and spectral resolution in the UV are superior to those in the X-rays, X-ray absorption spectra provide complete coverage of all ionization states, in contrast with UV absorption features which arise only from relatively low ionization gas. The underlying physical processes are absorption from  inner atomic (K- and L-) shells at X-ray energies. Most notable is Fe, in which the full range of charge states from neutral to H-like Fe$^{25+}$ can be (\citealt{Behar+01}), and is often (\citealt{Sako+01}) observed in X-ray absorption spectra of AGN. 

Defining the ionization parameter in terms of the ionizing luminosity $\lion$, gas density $\nH$, and distance from the ionizing source $r$:
\begin{equation}
\label{eq: xi}
\xi = \frac{\lion}{\nH r^2} ~~~,
\end{equation} 
Seyfert outflows can span up to five orders of magnitude in $\xi$ (\citealt{Steenbrugge+03, Steenbrugge+05, Costantini+07, Holczer+07}), namely $-1 < \log \xi < 4$, in cgs units. 
For practical modeling purposes, an AGN outflow is often crudely described in terms of several ionization and velocity components. 
UV spectra in particular typically resolve several outflow velocities (\citealt{Kraemer+01a,Kraemer+01b}), for the most part below 1000 km\,s$^{-1}$. 
Some X-ray spectra of Seyferts also exhibit more than one velocity component (\citealt{Detmers+11, HolczerBehar12}).
Common practice, thus, is to fit spectra with absorbing components in which the outflow velocities $v$ and the ionization parameters $\xi$ are discrete parameters of the model.
Indeed, the literature includes models with distinct high and low ionization components, as well as discrete high and low velocities, which fit X-ray spectra of Seyfert outflows fairly well (e.g., \citealt{Krongold+05, Steenbrugge+09, Kaastra+12}). 
Nonetheless, since in most cases the velocity range is small (often unresolved in X-rays), while the range in ionization is huge, we find it worthwhile to think of discrete velocity components, each having an internal ionization structure. 

In order to characterize the ionization structure of an outflow, \cite{Holczer+07} defined an absorption measure distribution (AMD), which is the distribution of column density $\Nh$ along the line of sight as a function of $\log \xi$:
\begin{equation}\label{eq: AMD definition}
 {\rm AMD} \equiv \left|\frac{\d\Nh}{\d \log \xi}\right|
\end{equation}
Reciprocally, the total column density $\Ntot$ can be expressed as an integral over the AMD. The AMD is the absorption analog of the emission measure distribution (EMD) widely used in the analysis of emission-line spectra. It provides a more complete representation of the ionization distribution than that of the more commonly used models of several ionization components each with a fixed $\xi$. 

\cite{Behar09} highlighted that Seyfert outflows exhibit a broad and flat AMD, which is remarkably similar between different objects.
A broad and flat ionization distribution is also suggested by the X-ray emission line spectrum of the Seyfert 2 galaxy NGC~1068, as analyzed by \citeauthor{Brinkman+02} (2002, fig.~9 therein) and \citeauthor{Kallman+14} (2014, fig.~11 therein). 
\cite{Behar09} showed that the broad AMD could either be due to a large-scale density profile in the flow of $n \propto r^{-1}$, such as in the MHD winds of \cite{Fukumura+10a, Fukumura+10b}, or alternatively, the broad AMD could be due to steep density gradients $n \propto x^{-1}$ inside a well-localized flow (e.g., cloud, where $x$ is the depth into the cloud). 
The most common interpretation of the outflow as a localized flow is the `thermal instability' model (\citealt{Krolik+81} and citations thereafter). 
In this model, the gas pressure $\pgas$ ($=2.3\nH k T$, where $T$ is the gas temperature) is such that a stable thermal equilibrium solution can be reached with more than one value of $T$. The different solutions, or `phases', are assumed to co-exist in pressure equilibrium, each with its own $T$ and $\nH$. 

However, as we show below, the values of $\xi$ observed in Seyfert outflows imply that the radiation pressure of the incident radiation $\prad$ ($=\lion/4\pi r^2 c$)\footnote{We emphasize that the definition of $\prad$ includes only the directed pressure of the incident AGN radiation, without any contribution from radiation which is emitted by the gas. Also, the value of $\prad$ equals the radiation pressure at the illuminated surface, before any absorption or scattering.} is much larger than $\pgas$. Therefore, if the gas is not accelerating, the slab must develop a gradient in $\pgas$ to counteract the force of radiation, in contrast with the uniform $\pgas$ assumed by the `thermal instability' model. This gas pressure gradient implies a sharp density gradient within a single localized slab. This local density gradient is inevitable, and is a general property of photoionized gas. 

This hydrostatic compressing effect of radiation pressure has been incorporated in the Seyfert outflow models of \cite{Rozanska+06} and \cite{Goncalves+07}, in models of \hii\ gas in star-forming regions (\citealt{Pellegrini+07,Draine11a,Verdolini+13,Yeh+13}), and in models of AGN emission line clouds by \cite{Dopita+02}, \cite{Groves+04}, \citeauthor{Stern+14} (2014, hereafter Paper I) and \citeauthor{Baskin+14} (2014, hereafter Paper II). 
\citeauthor{Rozanska+06}, \citeauthor{Goncalves+07}, and Papers I and II showed that a slab compressed by radiation pressure indeed spans a very broad range of ionization states, as commonly observed in Seyfert outflows. 

In this paper, we further show that radiation pressure compression produces a universal AMD, which is only weakly dependent on model parameters, and is insensitive to the uncertainties in radiation transfer calculations. We show that the predicted AMD is consistent with the observed AMD, and explains the small dispersion of the AMD between different objects. 
We emphasize that we do not attempt to produce a full kinematic model of Seyfert outflows, but rather show that independent of the exact kinematic model, the AMD has a universal shape. In the next Paper in this series (Baskin et al., in prep., hereafter Paper IV), we apply radiation pressure compression also to Broad Absorption Line Quasars.

The Paper is built as follows. 
In Section 2, we derive the AMD implied by radiation pressure compression, both by analytic approximation and by using full numerical radiative transfer calculations. 
In Section 3 we compare the derived AMD with available observations of Seyfert outflows. 
We discuss our results and their implications in Section 4, and conclude in Section 5.

\section{Theory}\label{sec: theory}

\newcommand{\tautot}{\taubar_{\rm tot}}
\subsection{RPC -- Conditions}\label{sec: rpc definition}
Assume a body of gas, irradiated by an ionizing source far enough relative to its size so that the incoming radiation is plane-parallel. 
Additionally, assume that any processes on the dimensions perpendicular to the radiation have long enough timescales so that they can be disregarded, and the analysis can be done in one dimension.
Also, assume that the gas can be treated as hydrostatic in its rest-frame, i.e. it is not significantly accelerated by the radiation. 
Now, if radiation is the dominant force applied to the gas in its rest frame, then the hydrostatic equation takes the form
\begin{equation}\label{eq: hydrostatic only radiation}
\frac{\d\pgas}{\d x} = \frac{\frad(x)}{c}\nH\sigbar(x) ~~~,
\end{equation}
where $\frad(x)$ is the flux of the ionizing radiation ($1-1000\ryd$) at a certain depth $x$, and $\sigbar$ is the spectrum-averaged radiation pressure cross section per H-nucleon
\begin{equation}\label{eq: sigbar definition}
 \sigbar\equiv\frac{\int_1^{1000\ryd}  \left( \sigma_{\rm abs}(\nu) + \sigma_{\rm sca}(\nu) \right) F_\nu\d\nu }{\int_1^{1000\ryd}F_\nu\d\nu } ~~~,
\end{equation}
where $F_\nu$ is the flux density, and $\sigma_{\rm abs}$ and $\sigma_{\rm sca}$ are the absorption and electron scattering cross-section per H-nucleon, respectively.
In the definition of $\sigbar$ we have disregarded a correction factor due to anisotropic scattering (see, e.g., \citealt{Draine11b}), which is of order unity. 
This simplification facilitates the use of $\sigbar$ also in the definition of the flux-averaged optical depth:
\begin{equation}
\d\taubar = \nH\sigbar \d x ~~~.
\end{equation}
Using the definitions of $\taubar$ and $\prad$ ($=\frad(0)/c$), we can replace $\frad(x)/c$ in eq. \ref{eq: hydrostatic only radiation} with $\prad e^{-\taubar}$. Therefore, the hydrostatic equation obtains the simple form 
\begin{equation}\label{eq: hydro simple form}
 \d\pgas(\taubar) = \prad e^{-\taubar}\d\taubar ~~~,
\end{equation}
with the solution 
\begin{equation}\label{eq: hydro solution}
 \pgas(\taubar) = \pgasz + \prad\left(1-e^{-\taubar}\right) ~~~,
\end{equation}
where a subscript $0$ denotes the value of a quantity at the illuminated surface.

Eq. \ref{eq: hydrostatic only radiation} neglects non-thermal sources of isotropic pressure, such as magnetic pressure $\pmag$ and the pressure of the trapped line emission $\pline$\footnote{Note that the directed pressure induced by line absorption enters $\sigma_{\rm abs}(\nu)$ above.}. These terms are assumed to be small compared to $\pgas$. We address the effect of relaxing this assumption in the appendix. 
Also, the right hand side of eq. \ref{eq: hydrostatic only radiation} should include a correction factor due to the pressure induced by non-ionizing photons. However, in dustless gas, as is probably the case in ionized AGN outflows (see below), the correction factor is of order unity and therefore we use eq. \ref{eq: hydrostatic only radiation} as is. This correction factor, and the effect of anisotropic scattering, are incorporated in the numerical calculations below.

We define a slab as Radiation Pressure Compressed (RPC)\footnote{Equivalently the gas may be described as Radiation Pressure Confined, as used in Papers I and II for the RPC acronym.}, if $\pgas$ increases significantly throughout the slab, i.e. $\pgas(\tautot) \gg\pgasz$, where $\tautot$ is the total optical depth of the slab. 
Equation \ref{eq: hydro solution} shows that $\prad\gg\pgasz$ is a necessary condition for RPC. How does this condition relate to the ionization state of the gas? Using the definition of $\xi$ (eq. \ref{eq: xi}), the condition
\begin{equation}
\frac{\prad}{\pgasz} = \frac{\lion / \left(4\pi r^2 c\right)}{2.3 n_{{\rm H},0} k T_0} \gg 1  ~~~,
\label{eq: cond1}
\end{equation}
is equivalent to 
\begin{equation}\label{eq: min xi}
 \xi_0 \gg 1.2\,\frac{T_0}{10^4\K} ~~~, 
\end{equation}
where for mathematical brevity, we implicitly assume that the numerical value of $\xi$ is in cgs units. 
Now, $\xi/T$ is a monotonically increasing function of $\xi$ (see below). Therefore, the fact that Seyfert outflows commonly exhibit gas with $\xi\gg1$ indicates that $\prad \gg \pgasz$ is a common property of Seyfert outflows. 

For a slab with $\tautot\gg 1$ to be RPC, the $\prad\gg\pgasz$ condition is sufficient, as noted by \cite{Dopita+02}, and Papers I and II. However, in slabs with $\tautot\lesssim 1$, such as ionized AGN outflows, $\pgas(\tautot) \approx \pgasz+\prad\tautot$, and therefore RPC also requires a minimum optical depth of 
\begin{equation}
\tautot \gg \frac{\pgasz}{\prad} ~~~
\label{eq: cond2}
\end{equation}
or equivalently
\begin{equation}
\tautot \gg 1.2\,\frac{T_0}{10^4\K}~\xi_0^{-1} ~~~.
\label{eq: tautot}
\end{equation}
Equation \ref{eq: tautot} emphasizes that in gas with $\xi\gg1$, absorption of only a small fraction of the radiation is sufficient to compress the gas, indicating that it is likely that Seyfert outflows are RPC. 

\begin{figure}
\includegraphics{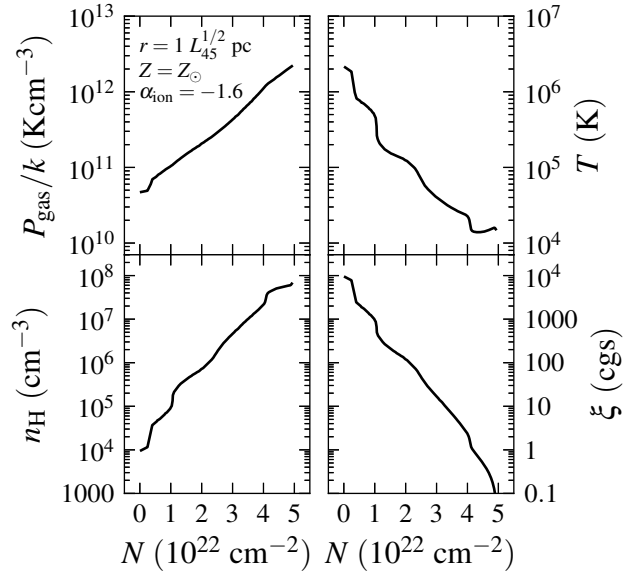}
\caption{
Pressure, temperature, density and ionization structure of a single RPC slab, as calculated with \cloudy.
Model parameters are noted in the top-left panel. 
The compression of the slab due to the absorption of radiation momentum causes $\pgas$ to increase by a factor of $50$ (top-left panel) from the illuminated surface ($N=0$) to the back side of the slab ($N=5\times10^{22}\cm^{-2}$). 
The increase in $\pgas$ is accompanied by a factor of 140 decrease in $T$ (top-right panel). 
The combined increase in $\pgas$ and decrease in $T$ implies a factor of $7000$ increase in $\nH$ (bottom-left panel), and hence a large dynamical range in $\xi$ (bottom-right panel). The implied AMD (eq. \ref{eq: AMD definition}) of the RPC slab is the slope of the $N$ vs. $\log \xi$ relation. 
}
\label{fig: RPCbasic}
\end{figure}

\subsection{RPC -- Slab structure}

What is the $\pgas$ and $\xi$ structure of an optically thin RPC slab?
The value of $\pgas$ on the back side of an RPC slab can be approximated as 
\begin{equation}\label{eq: pgas leading edge}
\pgas(\tautot) = \pgasz + \prad\tautot \approx \prad\tautot = \frac{\lion}{4\pi r^2 c}\tautot ~~~.
\end{equation}
To derive the full $\pgas$ and $\xi$ structure of an RPC slab we use \cloudy\ (version 13.03, \citealt{Ferland+13a}). 
\cloudy\ divides the slab into `zones', and solves the local thermal equilibrium and local ionization equilibrium equations for a given $\pgas$ in each zone, where each consecutive zone is subject to the attenuated radiation from the previous zone. We use the `constant pressure' flag in \cloudy\ (\citealt{Pellegrini+07}), which implies that \cloudy\ calculates $\pgas$ in each zone from the condition of hydrostatic equilibrium (eq. \ref{eq: hydrostatic only radiation}). Also, we turn off the pressure induced by the trapped emitted radiation in the calculation, and refer to the effect of this pressure term in the appendix. 

We note that there are several other available codes capable of solving the ionization state and slab structure of ionized AGN outflows. 
In the appendix we compare the \cloudy\ results with an analysis using \xstar\ (\citealt{KallmanBautista01}). 
\cite{Rozanska+06} used \titan\ (\citealt{Dumont+00}), in which the slab solving scheme enables selecting between different possible $T$ in zones where the solution to the thermal equilibrium equation is multi-valued, in contrast with the scheme in \cloudy\ which leads to a single solution only. 
A comparison with the analysis of \citeauthor{Rozanska+06} appears in the discussion. 

The input parameters of the \cloudy\ calculation are as follows. At $\hnu<1\ryd$ we use the standard AGN SED described in Paper I. The spectral slope of the ionizing radiation at $1\ryd-2\kev$ is parametrized by $\aion~(L_\nu \propto \nu^{\aion})$. We assume a power law with index --1 at $2-200\kev$ (\citealt{Tueller+08}; \citealt{Molina+09}) and a cutoff at larger frequencies. We use the default solar composition in \cloudy, with the abundances scaled with the metallicity parameter $Z$. The non-linear scaling of Helium and Nitrogen are detailed in Paper I. Models in this section are dust-free, the effect of dust grains is addressed below.

Figure \ref{fig: RPCbasic} plots an example of the structure of a RPC slab versus $N$, the Hydrogen column density measured from the illuminated surface. 
We assume an ionization parameter at the slab surface $\xi_0=10^4$, a total column density of $\Ntot=5\times10^{22}\cm^{-2}$, and the $r$, $\aion$ and $Z$ noted in the figure. The value of $r$ scales as $L_{45}^{1/2}$, where $\lion=10^{45}L_{45}\ergs$, since the models are mainly sensitive to $\frad$, rather than directly to the value of $r$. The dependence of the slab structure on these five parameters is discussed below. 
The top-left panel shows the most basic property of RPC. The value of $\pgas$ increases by a factor of $50$ from $N=0$ to $N=\Ntot$, due to the absorption of radiation momentum (eq. \ref{eq: hydro solution}). 
The top-right panel shows that the increase in $\pgas$ is accompanied by a drop in $T$, since generally $T$ decreases with decreasing $\prad/\pgas$ (see \citealt{Chakravorty+09} for a recent analysis of $T$ vs. $\prad/\pgas$). 
The combined increase in $\pgas$ and drop in $T$ implies an increase by a factor of $7\,000$ in $\nH$ (bottom-left panel). 

The lower-right panel shows the $\xi$ structure of the RPC slab, which spans a large dynamical range ($-1 < \log \xi < 4$), as previously noted by \cite{Rozanska+06}. 
We note that in the calculation of $\xi$, we take into account both the change in $\nH$ and the absorption of ionizing photons.
At $0<\Nh<4\times10^{22}\cm^{-2}$, where only $50\%$ of the ionizing energy is absorbed by the slab, $\xi$ drops from the assumed $\xi_0=10^4$ to $\xi=1.2$ due to the increase in $\nH$ seen in the bottom-left panel. This range of $\xi$, in layers which do not significantly absorb the radiation, is a unique property of RPC, and is in contrast with the constant $\xi$ expected in constant density models. 

The bottom-right panel of Fig. \ref{fig: RPCbasic} also suggests that the AMD in a RPC slab is flat, since $N$ vs. $\log \xi$ has a roughly constant slope, i.e. a constant $\left|\d N/\d \log \xi\right|$, or a flat AMD. In the following sections we examine this flatness of the AMD, and its dependence on model parameters. 
A flat AMD is a result of a density profile of the form $\nH \propto x^{-1}$ (\citealt{Behar09}; the physical depth into the cloud $x$ was denoted there by $\delta r$). The $\nH\propto x^{-1}$ profile is evident in the bottom-left panel of Fig. \ref{fig: RPCbasic}, which shows that $\d\log\nH \propto \d\Nh$, implying that
\begin{equation}
 \frac{\d\log\nH}{\d x} \propto \frac{\d\Nh}{\d x} = \nH ~~~,
\end{equation}
hence
$\d\nH/\d x \propto \nH^2$, the solution of which is $\nH\propto x^{-1}$. 

We are not aware of any other physical mechanism, which produces a similar $\nH \propto x^{-1}$ structure. In particular, turbulence in the interstellar medium (ISM) also creates density gradients over a wide range of densities. However, its typical density variations are quite different than $\nH \propto x^{-1}$. ISM variations generally follow the \cite{Kolmogorov41} power spectrum for non-magnetic incompressible turbulence of $P(k) \propto k^p$ with $p = 11/3$, which implies a density scaling of $\nH \propto x^{(p-3)/2} = x^{1/3}$, which indeed is quite different from the RPC scaling of $\nH \propto x^{-1}$. For more details see \cite{Behar09}.

\subsection{RPC -- The Absorption Measure Distribution}

What is the AMD expected in RPC gas?
We first present the numerical calculation of the AMD by \cloudy. Then, in order to gain physical insight on the properties of the calculated AMD, we present an approximate analytic solution.

\begin{figure*}
\includegraphics{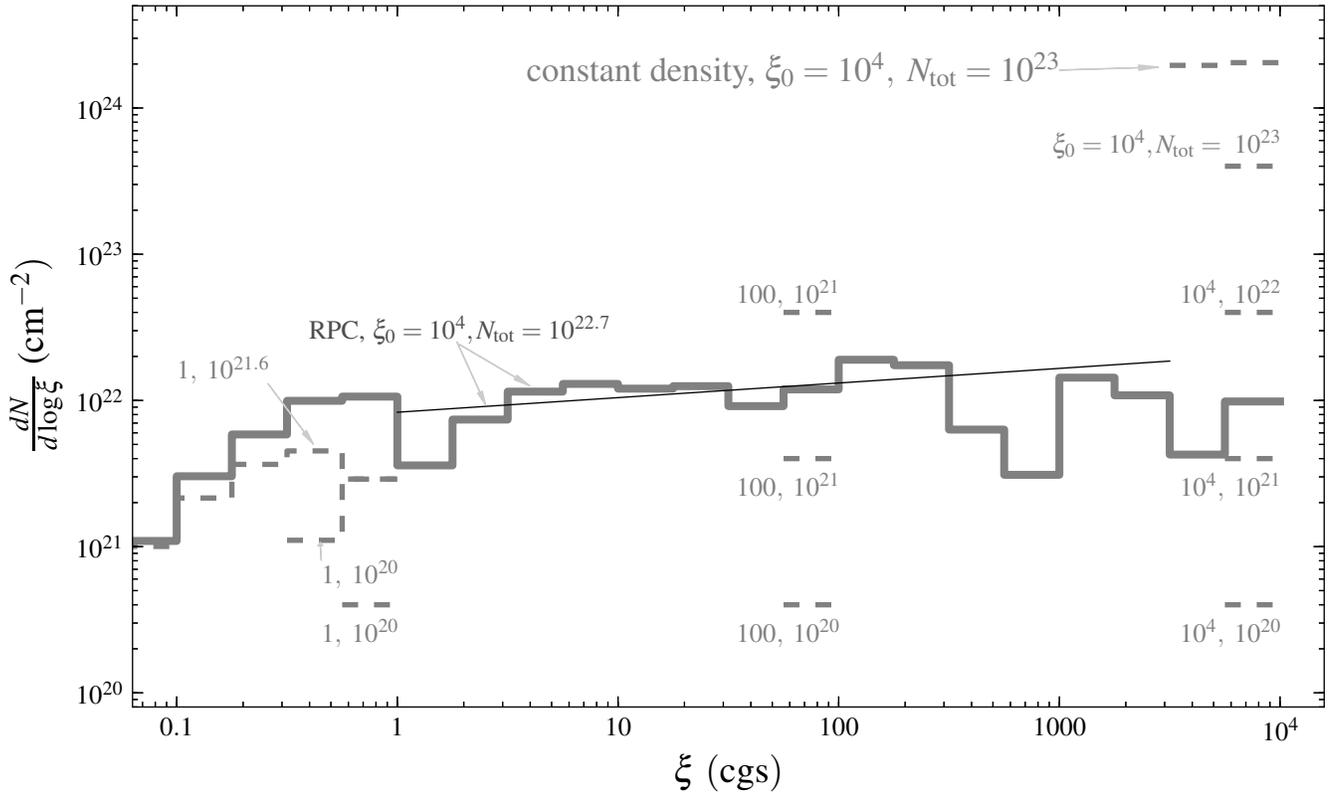}
\caption{
The Absorption Measure Distribution of the RPC slab shown in Fig. \ref{fig: RPCbasic} (thick solid line) vs. that of 12 constant density slabs (dashed lines),
with different $\xi_0$ and $\Ntot$. 
AMDs are derived from the \cloudy\ solution of the slab structure, using bins of 0.25 dex in $\xi$. 
All models have $\tau(0.5-1.5\kev)<1$, as required by observations. 
The AMD in the RPC model is flat, and spans a large range of $\xi$. 
Conversely, most constant density models have a single $\xi$ throughout the slab. 
Only models with sufficient $\Ntot$ to absorb a significant fraction of the ionizing radiation span a (small) range of $\xi$.
In order to reproduce an AMD with a large range of $\xi$ with constant density models, an ad-hoc tiling of several slabs with different $\xi$ and specific $\Ntot$ would be required. 
The thin solid line plots eq. \ref{eq: AMD estimate}, a semi-analytic approximation of the AMD at $1<\xi<3000$ in the RPC model. 
}
\label{fig: AMD vs const density}
\end{figure*}

\subsubsection{Numerical evaluation}
To derive the AMD from the \cloudy\ calculation, we subsample the \cloudy\ zones into bins of 0.25 dex in $\xi$ (each zone has a single $n$ and therefore a single $\xi$), and calculate the AMD $=\Delta\Nh/0.25$ by summing $\Delta N$ of all zones in the same $\xi$-bin. This sampling is sufficient since it has higher resolution than the resolution of $0.5-1$ dex achievable by current observations (see below). 
Figure \ref{fig: AMD vs const density} plots the implied AMD for the RPC model shown in Fig. \ref{fig: RPCbasic}. 
Note that at $\xi<1$ the absorption of the ionizing continuum is significant, and therefore layers with $\xi<1$ in a RPC slab are not entirely equivalent to slabs with the same $\xi$ at their illuminated surface, due to the difference in the incident spectrum. 

The AMD in the RPC model is flat, with a value remaining in the range of $10^{21.5}-10^{22.3}\cm^{-2}$ over a factor of $10^5$ in $\xi$. This flatness can be quantified by approximating the calculated AMD as a power-law, $\d N/\d \log \xi\approx N_1\xi^a$. The best fit power law is
\begin{equation}\label{eq: AMD numeric}
 \left.\frac{\d N}{\d \log \xi}\right|_{\rm numeric} = 7.6\times10^{21} \xi^{0.03} \cm^{-2} ~~~.
\end{equation}

\subsubsection{Analytic approximation}\label{sec: analytical}

The differential equation for the AMD can be derived from the equation of hydrostatic equilibrium (eq. \ref{eq: hydro simple form}).
It is instructive to introduce the pressure ionization parameter, $\Xi\equiv\prad/\pgas=\xi/9.2\pi c k T$.\footnote{We define $\Xi$ as in \cite{Dopita+02} and in Paper I, which implies a factor of 2.3 smaller value than the definition in \cite{Krolik+81}.}
In these units, the RPC conditions (eqs. \ref{eq: cond1} and \ref{eq: cond2}) have the simple form 
\begin{equation}\label{eq: boundary Xi}
 \Xi_0 \gg 1 ~~~, \\
\end{equation}
and
\begin{equation}
 \tautot \gg \Xi_0^{-1}~~~.
\end{equation}
On the illuminated side of the absorber, where the optically thin limit $e^{-\taubar}\sim 1$ holds, eq. \ref{eq: hydro simple form} can be expressed as 
\begin{equation}
 \d\left(\frac{1}{\Xi}\right) = \d\taubar ~~~.
\end{equation}
With some algebra, we get
\begin{equation}\label{eq: rpc prediction}
\left|\frac{\d \taubar}{\d \log \Xi}\right| = \Xi^{-1} ~~~.
\end{equation}
Since $\left|\d\taubar/\d\log\Xi\right|$ is the fraction of the energy and momentum absorbed at each decade of $\Xi$, and in RPC gas $\Xi_0\gg1$ (eq. \ref{eq: boundary Xi}), eq. \ref{eq: rpc prediction} implies that the gas layer with $\log\ \Xi\sim 2$, for example,  will absorb $1\%$ of the radiation, the layer with $\log \Xi\sim 1$ will absorb $10\%$ of the radiation, and a significant fraction of the radiation will only be absorbed at $\log \Xi \sim 0$, i.e. when $\pgas\approx\prad$. 
 
The AMD, $\left|\d N/\d\log\xi\right|$, is closely related to $\left|\d \taubar / \d \log \Xi\right|$. Replacing $\d\taubar$ with $\sigbar\d\Nh$ and $\Xi$ with $\xi / (9.2\pi c \kB T)$ in eq. \ref{eq: rpc prediction} we get 
\begin{equation}
 \left|\frac{\sigbar \d \Nh}{\d \log \frac{\xi}{T}}\right| = 9.2\pi c \kB\frac{T}{\xi} ~~~.
\end{equation}
Now, since
\begin{equation}
 \d \log \frac{\xi}{T} = \left(1 - \frac{\d \log T}{\d \log \xi}\right)\d \log \xi  ~~~,
\end{equation}
we get that in RPC
\begin{eqnarray}\label{eq: AMD}
 AMD &\equiv& \left|\frac{\d\Nh}{\d \log \xi}\right| = 9.2\pi c \kB\frac{T\left(1 - \frac{\d \log T}{\d \log \xi}\right)}{\sigbar\xi}  \nonumber \\
 & = &1.2 \cdot \frac{1}{\sigbar} \cdot \frac{T}{10^4\K}\cdot\xi^{-1} \cdot \left(1 - \frac{\d \log T}{\d \log \xi}\right) ~~\cm^{-2} ~~~.
\end{eqnarray}
Equation \ref{eq: AMD} shows that the AMD in RPC gas is inversely proportional to $\sigbar$, as expected. 
Also, it is proportional to $T/\xi$ $(\propto \Xi^{-1})$, because a smaller fraction of the radiation is absorbed at higher $\Xi$ (eq. \ref{eq: rpc prediction}). 
The last term comes from the factor of $T$ difference between $\xi$ and $\Xi$. The AMD will be small at $\xi$ where $T$ decreases rapidly with decreasing $\xi$. 

Is there a solution where $(1 - \d\log T / \d\log \xi) \leq 0$ and the AMD is formally zero or negative? 
The nature of such solutions can be understood by noting that $(1 - \d\log T/\d\log\xi)=(1+\d\log T/\d\log \Xi)^{-1}$. Since gas with $\d\log T/\d\log \Xi<0$ is thermally unstable (\citealt{Krolik+81}), then all solutions with $(1 - \d\log T / \d\log \xi) < 0$ are excluded. 
However, if $T$ has two stable values for a given $\Xi$, then a transition between the two branches of the solution implies a $T$-discontinuity, or $\d \log T/\d\log\Xi \rightarrow \infty$, which implies $1 - \d\log T/\d\log\xi\rightarrow 0$. Therefore, a $T$-discontinuity (sometimes called a `thermal front') will appear as a range of $\xi$ which has a null AMD.

\begin{figure*} 
\includegraphics{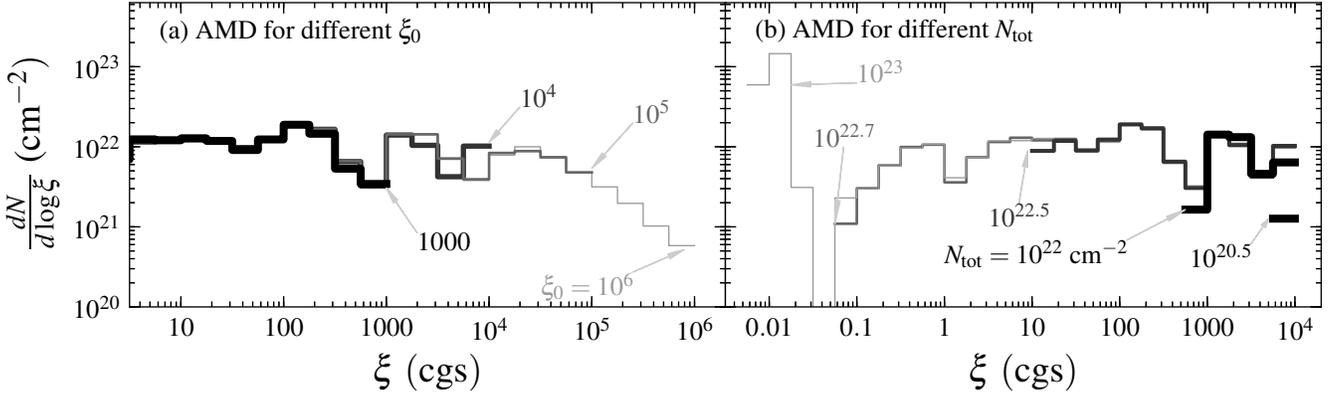}
\caption{
The dependence of the AMD on boundary values in RPC gas.
{\bf (a)}
\cloudy\ calculations of the AMD in models with different $\xi_0$.
All models lie on a similar ${\rm AMD}(\xi)$ solution, the models differ only in the `starting point'.
Hence, in RPC gas, $\xi_0$ determines the highest $\xi$ in the slab. However, the AMD at $\xi < \xi_0$ is independent of $\xi_0$.
The unobservable fully ionized layer $(\xi>10^{4.5})$, which exhibits a drop in the AMD with increasing $\xi$, is discussed in \S\ref{sec: fully ionized}.
{\bf (b)}
The AMD for different $\Ntot$. All models lie on a similar ${\rm AMD}(\xi)$ solution, differing only in the `end point'. 
}
\label{fig: AMD vs N and xi}
\end{figure*}

Equation \ref{eq: AMD} can be further simplified if $\sigbar$ and $T$ are approximated as power laws in $\xi$:
\begin{equation}\label{eq: powerlaws}
 T(\xi) = T(1) \xi^p ~, ~~~~~ 
 \sigbar(\xi) = \sigbar(1) \xi^{-q} ~~~,
\end{equation}
where $T(1)$ and $\sigbar(1)$ are the values of $T$ and $\sigbar$ at $\xi=1$, and $p, q$ are constants. In Appendix \ref{app: coeffs}, we address the validity of this approximation by comparing it with the full \cloudy\ numerical calculation of $\sigbar(\xi)$ and $T(\xi)$. 
Using eq. \ref{eq: powerlaws} in eq. \ref{eq: AMD}, we obtain a power-law form for the AMD:
\begin{eqnarray}\label{eq: AMD approx}
 \left|\frac{\d \Nh}{\d \log \xi}\right| &=& N_1 \xi^a \nonumber \\
 N_1 &=& 1.2\frac{1}{\sigbar(1)} \frac{T(1)}{10^4\K} (1 - p) \cm^{-2} \nonumber \\
 a &=& p + q - 1 
\end{eqnarray}
Eq. \ref{eq: AMD approx} implies that if $p+q\approx1$, then the AMD will be flat. 

One can obtain a rough estimate of $p$ by noting that $T$ equals the Compton temperature $\tc$ ($=4\times10^6\K$ for our assumed SED) when the gas is fully ionized (i.e. at $\xi\gtrsim10^4$), while $T\approx10^4\K$ near the transition between ionized and neutral gas (i.e. at $\xi\lesssim 1$), suggesting that $p \sim (6.6-4) / (4 - 0) = 0.65$. 
On the other hand, $q$ depends on the source of opacity. If the opacity is dominated by a single bound-free edge, then photoionization equilibrium in ionized gas implies $\sigbar \nH \propto \nH^2$, and hence $\sigbar \propto \nH \propto \xi^{-1}$, i.e. $q=1$. In contrast, if electron scattering dominates the opacity, $\sigbar$ is independent of $\xi$ and therefore $q=0$. In practice, the dominant source of absorption is \hi\ at low $\xi$, metal edges at intermediate $\xi$, and electron scattering at high $\xi$ where the gas is fully ionized. Therefore, we expect $q$ to have some intermediate value between $0$ and $1$. 

In appendix \ref{app: coeffs} we show that the \cloudy\ calculation of the model shown in Figs. \ref{fig: RPCbasic}--\ref{fig: AMD vs const density} suggests that $T(\xi)$ and $\sigbar(\xi)$ can be approximated as power-laws between $1<\xi<3000$, with $T(1)=10^{4.2}\K$, $\sigbar(1)=10^{-22}\cm^2$, $p=0.5$ and $q=0.6$. 
Using these values in eq. \ref{eq: AMD approx}, we find
\begin{equation}\label{eq: AMD estimate}
 \left. \frac{\d \Nh}{\d \log \xi}\right|_{\rm semi-analytic} = 8.3\times10^{21} \xi^{0.1} \cm^{-2} ~~~,
\end{equation}
i.e. a flat AMD, since $p+q\approx1$. 
The expression in eq. \ref{eq: AMD estimate} is shown in Fig. \ref{fig: AMD vs const density}, spanning the range of $\xi$ on which it was derived. 
It can be seen that it is very close to the result of the detailed numerical calculation (see also eq. \ref{eq: AMD numeric}).  
At $\xi>300$, the analytic AMD somewhat overestimates the numerical calculation, because at these $\xi$ the opacity is dominated by electron scattering, and therefore non-ionizing radiation contributes significantly to the compression of the gas, a term which we neglected in the analytic approximation. 

\subsection{Comparison of the AMD in RPC with the AMD in constant density models}

Fig. \ref{fig: AMD vs const density} also shows the AMD of constant density models. 
The constant density models are calculated for a range of $\Ntot$ and $\xi_0$, under the requirement that the optical depth at $0.5-1.5\kev$ is $<1$, as implied by observations (\citealt{Reynolds97}). Other model parameters are identical to the RPC model. It can be seen that optically-thin constant-density models have a single $\xi$ throughout the slab, as expected since both $\nH$ and the ionizing flux do not change within the slab. 
Only models with sufficient $\Ntot$ to absorb a significant fraction of the ionizing radiation span some range of $\xi$ values. In constant density models with $\xi_0=10^4$, significant absorption occurs at $\Nh \geq 10^{24}\cm^{-2}$, in $\xi_0=100$ at $\Nh \geq 10^{23}\cm^{-2}$ (not shown, due to large $\tau$ at $1\kev$), and in $\xi_0=10$ at $\Nh \geq 10^{21}\cm^{-2}$. 
Therefore, a single constant-density model cannot span a large range of $\xi$ unless it is optically thick, and if it is optically thick, the AMD will be much steeper than observed. In order to reproduce the observed broad and flat AMD with constant density models, an ad-hoc tiling of several optically thin slabs with different $\xi$ and specific $\Ntot$ would be required. In contrast, a single optically thin RPC slab creates a broad and flat AMD.

\subsection{How does the AMD depend on model parameters?}\label{sec: AMD vs. params}

In this section, we show that in RPC, $N_1$ and $a$ only weakly depend on model parameters, implying that $N_1$ and $a$ are robust predictions of RPC.

\subsubsection{AMD versus $\xi_0$ and $\Ntot$}

Panel {\bf a} of Figure \ref{fig: AMD vs N and xi} shows the dependence of the AMD on $\xi_0$. It can be seen that the value of $\xi_0$ determines the maximum $\xi$ spanned by the AMD. However, at $\xi<\xi_0$, the AMD is practically independent of $\xi_0$. What causes this behavior? The model with $\xi_0=10^5$, for example, includes a layer with $10^3<\xi<10^5$ which does not exist in the $\xi_0=10^3$ model. Since this additional layer is optically thin, it does not significantly affect the solution at $\xi_0 < 10^3$. Therefore, the $\xi_0=10^3$ and $\xi_0=10^5$ models have similar solutions at $\xi<10^3$. 

Note the drop in the AMD at $\xi>10^{4.5}$. At these $\xi$, the gas is fully ionized (the dominant Iron ionization state is Fe$^{+26}$), and therefore it is unobservable in absorption, since it does not create discrete absorption features in the observed spectrum. We further discuss this fully ionized layer below. 

Panel {\bf b} of Fig. \ref{fig: AMD vs N and xi} shows the dependence of the AMD on $\Ntot$. 
The value of $\Ntot$ determines the minimum $\xi$ spanned by the AMD. A model with $\Ntot=10^{22}\cm^{-2}$ will have only $\xi>1000$ gas layers, while a model with $\Ntot=10^{22.7}\cm^{-2}$ will span $0.1<\xi<10^4$. 
However, the AMD at $\xi$ higher than the minimum $\xi$ is independent of $\Ntot$. 

Therefore, Fig. \ref{fig: AMD vs N and xi} implies that $\xi_0$ and $\Ntot$ have no effect on $N_1$ and on $a$. Rather, they only determine the starting point and end point on the universal AMD solution.

\begin{figure} 
\includegraphics{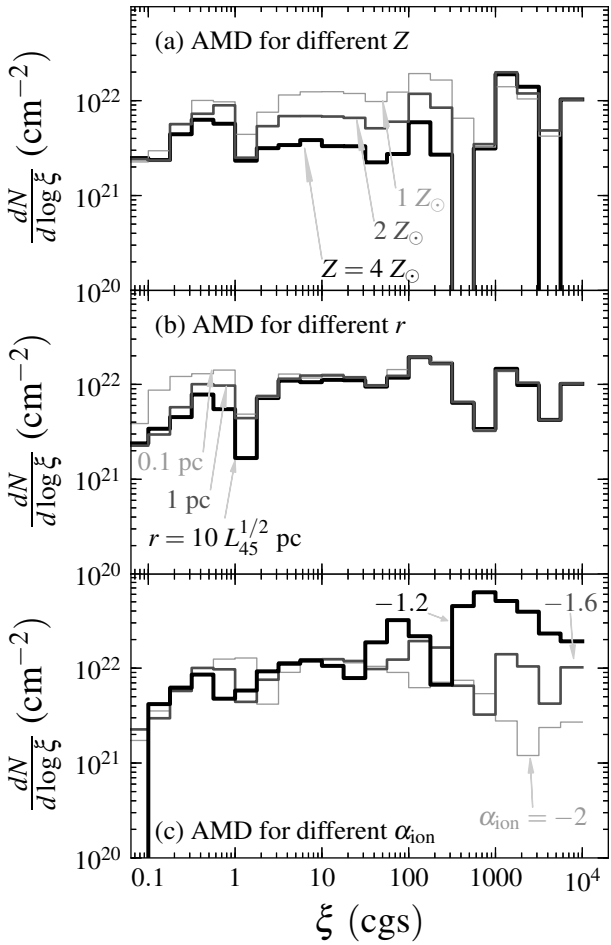}
\caption{The dependence of the AMD on model parameters in RPC gas.
{\bf (a)} 
\cloudy\ calculations of the AMD in models with different $Z$. The AMD is $\propto Z^{-1}$ at $1<\xi<300$, and independent of $Z$ at $\xi<1$. Models with $Z>1\zsun$ have a $T$-discontinuity at $\xi=400$, and the model with $Z=4\zsun$ has another discontinuity at $\xi=4000$.
{\bf (b)}
The AMD is only weakly dependent on $r$. 
{\bf (c)}
Models with flatter $\aion$ have generally higher AMDs at $\xi>30$. The AMD is practically independent of $\aion$ at $\xi<30$. 
}
\label{fig: AMD vs params}
\end{figure}

\subsubsection{AMD versus metallicity}
Panel {\bf a} in Figure \ref{fig: AMD vs params} shows the dependence of the AMD on $Z$, for $Z$ typical of the narrow line region in AGN (\citealt{SternLaor13}). The AMD scales as $Z^{-1}$ at $1<\xi<300$. This trend is related to the increase of $\sigbar$ with $Z$ at $\xi$ where metals dominate the opacity, which decreases the AMD (eq. \ref{eq: AMD}). At $\xi>300$, two $T$-discontinuities appear in the AMD with increasing $Z$. The dependence of the $T$-discontinuity on model parameters is briefly discussed in the next section. At $\xi<1$, where H and He dominate the opacity, the AMD is independent of $Z$.

\begin{figure*}
\includegraphics{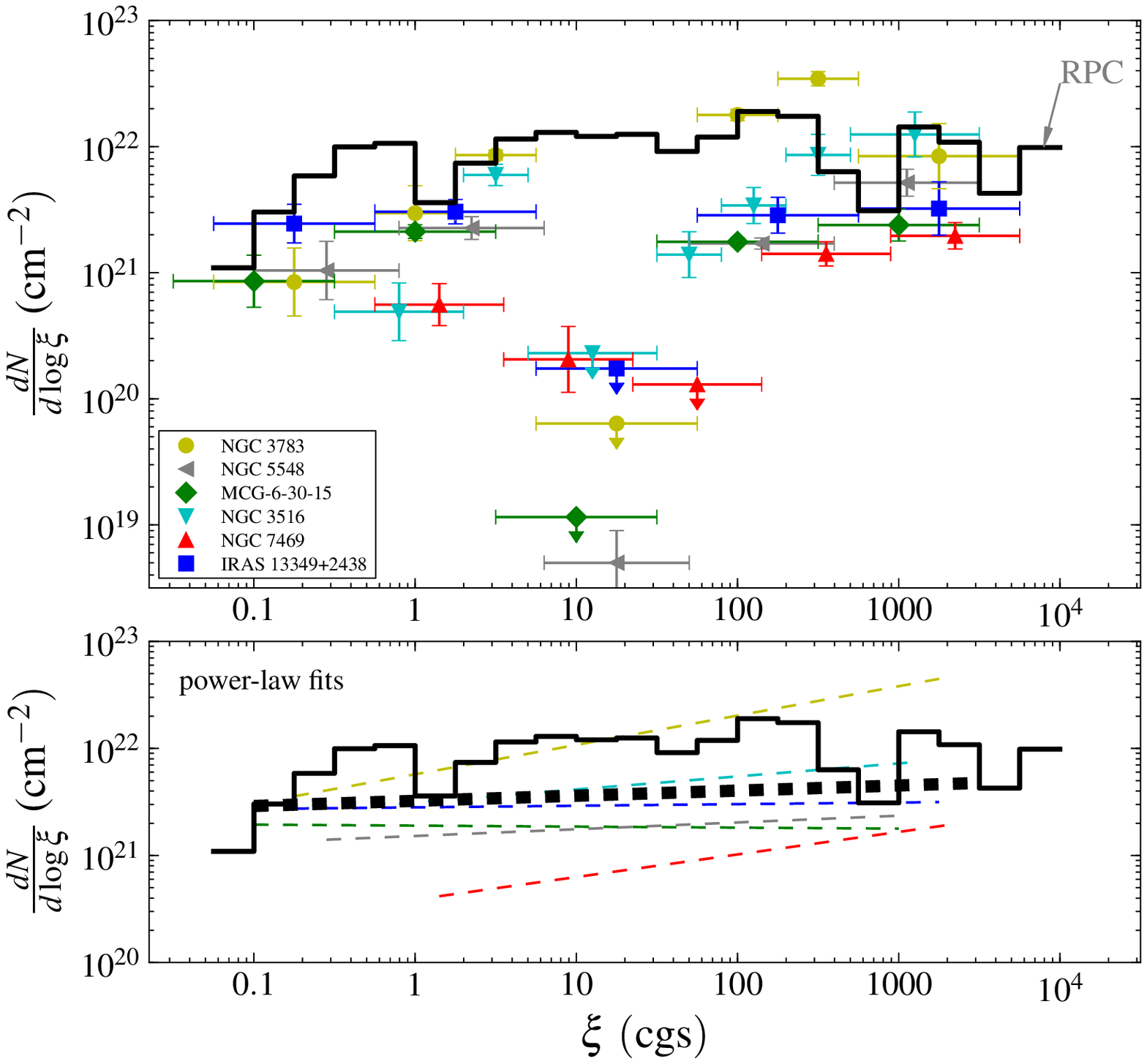}
\caption{
Absorption Measure Distribution of RPC gas, compared to observations.
Data points in the upper panel mark the measured AMD of six objects. 
The solid line denotes the \cloudy\ calculation of the AMD in a single RPC slab, from Fig. \ref{fig: AMD vs const density}. 
For clarity, dashed lines in the lower panel show power-law fits to the measured AMD data of each object, excluding upper limits. 
The thick black dashed line is the power-law fit to the measurements of all objects combined.
The average observed AMD slope is in excellent agreement with the  predicted slope, while the predicted AMD normalization is higher by a factor of $2.5$ than the observed average normalization. This general agreement between the AMD predicted by RPC and the observations, with no free parameters, strongly suggests that ionized AGN outflows are RPC. The \cloudy\ calculation does not predict the observed discontinuity at $\xi \approx 20$. 
}
\label{fig: AMD}
\end{figure*}

\subsubsection{AMD versus distance from the central source}\label{sec: AMD vs. r}

An upper limit on $r$ can be derived from the geometric requirement that the total slab depth is $<r$ (e.g. \citealt{Blustin+05}):
\begin{equation}\label{eq: Blustin}
 r > \frac{\Nh(\xi)}{n} = \frac{\Nh(\xi)}{\lion / \xi r^2}  ~~~.
\end{equation}
For example, NGC~3516 has $\approx10^{22}\cm^{-2}$ of $\xi\approx1000$ gas (see next section). 
Therefore, eq. \ref{eq: Blustin} implies
\begin{equation}\label{eq: r upper limit}
r < \frac{\lion}{\xi\Nh} = \frac{10^{45}L_{45}}{10^3\cdot10^{22}} = 30~L_{45}\pc ~~~.
\end{equation}
In RPC, all $\xi$-states come from the same slab, i.e., they have the same $r$, suggesting that the upper limit derived in eq. \ref{eq: r upper limit} applies to the gas at all $\xi$. 

Panel {\bf b} in Fig. \ref{fig: AMD vs params} shows the dependence of the AMD on $r$, for $r$ below the upper limit derived in eq. \ref{eq: r upper limit}. 
For a given $\xi$, $\nH\propto r^{-2}$, therefore the dependence of $\d\Nh/\d\log\xi$ on $r$ is effectively a probe of the dependence of $\d\Nh/\d\log\xi$ on $\nH$. It can be seen that the AMD is almost independent of the choice of $r$. 
We note that at $r$ smaller by several orders of magnitude than the values shown in Fig. \ref{fig: AMD vs params}, much higher $\nH$ would be required in order to obtain the observed $\xi$, and free-free absorption would become significant (\citealt{Rozanska+08}), which would in turn affect the AMD. 
However, the narrow line profiles observed in X-ray outflows suggest the gas is farther out than this high-$\nH$ regime, i.e. the outflows are in the low-$\nH$ regime, where the AMD is independent of $r$.

\subsubsection{AMD versus $\aion$}
Panel {\bf c} in Fig. \ref{fig: AMD vs params} shows the dependence of the AMD on $\aion$ for the range of slopes expected in AGN (see Paper II). Flatter $\aion$ generally have higher AMDs at $\xi>30$, due to the higher $T$ at a given $\xi$ implied by a flatter spectrum (see eq. \ref{eq: AMD}). The AMD is practically independent of $\aion$ at $\xi<30$. 

\subsection{The fully ionized layer $(\xi>10^{4.5})$}\label{sec: fully ionized}

The left panel of Fig. \ref{fig: AMD vs N and xi} shows the expected AMD at $\xi>10^{4.5}$, where all elements up to Fe are fully ionized and $T=\tc$. 
In this layer, momentum is transferred from the radiation to the gas via electron scattering, i.e. $\sigbar$ equals the electron scattering opacity. Since $T$ and $\sigbar$ are independent of $\xi$, the $p$ and $q$ defined in eq. \ref{eq: powerlaws} are zero, and hence eq. \ref{eq: AMD approx} implies that the ${\rm AMD}$ drops as $\xi^{-1}$, as seen in the left panel of Fig. \ref{fig: AMD vs N and xi}. 

Eq. \ref{eq: AMD approx} can also be used to derive the column density of this fully ionized layer\footnote{Including geometric dilution of the radiation with $r$ in the analysis can only decrease the upper limit on $\Nh_{\xi>10^{4.5}}$.}:
\begin{eqnarray}\label{eq: N large xi}
 \Nh_{\xi>10^{4.5}} &=& \int_{4.5}^{\log \xi_0} \frac{\d \Nh}{\d \log \xi}\d \log \xi \nonumber \\
&=& \int_{4.5}^{\log \xi_0} 2\times10^{22}\frac{\tc}{10^{6.6}\K}\left(\frac{\xi}{10^{4.5}}\right)^{-1}\d\log \xi  \nonumber\\
&\xrightarrow{\xi_0\rightarrow\infty} &0.8\times10^{22}\frac{\tc}{10^{6.6}\K} \cm^{-2} ~~~.
\end{eqnarray}
Since this fully ionized layer is unobservable, eq. \ref{eq: N large xi} implies that current estimates of $\Ntot$ in Seyfert outflows could be underestimated by up to $10^{22}\cm^{-2}$. 

Note that eq. \ref{eq: N large xi} suggests that electron scattering alters the state of the gas already at $N=10^{22}\cm^{-2}$, where the optical depth is merely $\approx 0.01$. This non-intuitive value originates from the fact that fully ionized gas must have $\xi>10^{4.5}$, which for $\tc=4\times10^6\K$ is equivalent to $\pgas<0.01\,\prad$. Hence, 1\% of the radiation momentum is sufficient to compress the gas so that it will not be fully ionized.

\section{Comparison with observed AMD}\label{sec: comparison}

In Figure \ref{fig: AMD} we compare the AMD expected in RPC with available observations. 
The error-bars in the top panel denote all AMDs currently measured, the five objects from \cite{Behar09} and NGC~3516 from \cite{HolczerBehar12}. In objects where the absorption has several components with different velocities $v$ (MCG-6-30-15 and NGC~3516), we plot the AMD of the slowest component (smallest $|v|$), and discuss faster components below. 
For details about the AMD formalism and how it is measured, see \cite{Holczer+07}. We emphasize here that the AMD is measured from column densities of Fe ions, and therefore the conversion to the H-column density shown in Fig. \ref{fig: AMD} depends on the assumed Fe/H. The observed AMDs were derived using the \cite{Asplund+09} abundances, which have Fe/H$=10^{-4.5}$, similar to the Fe/H$=10^{-4.55}$ used in the $Z=\zsun$ \cloudy\ models.

The observed AMDs have several common features described in \cite{Holczer+07} and in \cite{Behar09}. 
Most AMDs exhibit a bin consistent with no column at $\xi\approx20$, with NGC~7469 exhibiting a drop at somewhat higher values of $\xi=55$. 
Except the null bin, the AMDs are broad and flat, with $\d\Nh/\d\log\xi$ spanning a range of only $0.1-1.6$ dex over a range of $10^5$ in $\xi$. \cite{Behar09} quantified this behavior by fitting the observations with power-laws, ${\rm AMD}\propto N_1\xi^a$, excluding the drop (see table 1 there). The AMD of NGC~3516, which was measured later, is best-fit with $N_1=3\times10^{21}\cm^{-2}$ and $a=0.12$. These fits are shown as dashed lines in the bottom panel of Fig. \ref{fig: AMD}. The AMDs are remarkably similar, with all fit power-laws slopes in the range $0<a<0.4$. Also, the dispersion in the value of $\d N/\d\log\xi$ between different objects is $<0.4\,$dex in all $\log\xi$ bins. Note however that a larger and more well-defined sample is required to derive a more robust estimate of these dispersions. 

Fig. \ref{fig: AMD} also plots the AMD derived by \cloudy\ shown in Fig. \ref{fig: AMD vs const density}. 
Excluding the drop, the predicted AMD agrees with most of the observations to within a factor of three to five, where the predicted AMD tends to higher values than the observations. We note that despite the systematic offset between the prediction and the observations, the weak dependence of the predicted $N_1$ and $a$ on model parameters (\S\ref{sec: AMD vs. params}) is consistent with the small dispersion in the observations.

The bottom panel also shows the power-law fit to all objects combined (thick dashed line), which is 
\begin{equation}\label{eq: AMD observed}
  \left.\frac{\d N}{\d\log\xi}\right|_{\rm observed}=3\times10^{21}\xi^{0.05}\cm^{-2} ~~~.
\end{equation}
The average observed $a$ is in excellent agreement with the value of $a$ calculated by \cloudy, while the predicted $N_1$ is higher by a factor of $2.5$ than the average observed $N_1$ (compare eq. \ref{eq: AMD observed} with eq.  \ref{eq: AMD numeric}). 
This discrepancy between the observed and calculated AMD normalization is comparable to the discrepancy between the AMD calculated by \cloudy\ and \xstar\ (appendix \ref{app: xstar}), which is probably either due to differences in the metal lines used by the two codes, or due to differences in the recombination rate coefficients. Furthermore, \cite{Goosmann+11} compared the Fe column densities in an RPC slab calculated with \titan\ to observations of NGC~3783, and do not find a systematic offset, further supporting the possibility that the $N_1$ offset is due to a difference in the radiation transfer codes.

The fact that RPC reproduces the typical observed $a$, and the typical observed $N_1$ to within a factor of three, without any free parameters, strongly suggests that ionized AGN outflows are RPC. This is the main conclusion of this paper.

\subsection{The $T$-discontinuity}

The observations show a $T$-discontinuity centered at $\xi\approx20$, which span a decade in $\xi$, and corresponds to $T=10^{4.5}-10^{5}\K$. 
This discontinuity is clearly absent from the \cloudy\ calculation. Much narrower $T$ discontinuities are predicted at $\xi=400$ and $\xi=4000$ in the $Z=2\zsun$ and $Z=4\zsun$ models shown in Fig. \ref{fig: AMD vs params}. These discontinuities are actually also present in the $Z=\zsun$ model (see Fig. \ref{fig: RPCbasic}, at $N=0.3\times10^{22}\cm^{-2}$ and $N=1.1\times10^{22}\cm^{-2}$), though their width in $\xi$ is below our chosen resolution of $0.25\,$dex, and therefore they are not apparent in Fig. \ref{fig: AMD vs params}. All the predicted discontinuities are below the $\xi$ sensitivity of the observations, and thus even if they exist, are unlikely to be observed. However, why is the observed decade-wide discontinuity at $\xi\approx20$ not reproduced by \cloudy?

The $T$-discontinuity is usually associated with a thermal instability (\citealt{Holczer+07}).  
\cite{Hess+97}, \cite{Chakravorty+08, Chakravorty+09, Chakravorty+12}, and \cite{Ferland+13b} showed that the existence and properties of the thermal instability are highly sensitive to the assumed model parameters, and to the details of the atomic physics used in the code. Specifically, \cite{Hess+97} showed the sensitivity of the instability to the Fe abundance, and \cite{Chakravorty+12} showed the sensitivity of the instability to the exact shape of the ionizing continuum, which here we simplified as a single power-law with index $\aion$. 
Moreover, \citeauthor{Goosmann+11} (2011, see also Goosmann et al., in prep.), do obtain an unstable region around $T\approx10^5\K$ using a more complex incident spectrum between $1\ryd$ and $2\kev$, and a different radiation code (\titan). They also obtain instabilities at higher $T$, perhaps similar to those found by \cloudy. 
Therefore, the $T$-discontinuities are not robust predictions of the \cloudy\ calculation. We refer the reader to the mentioned papers and to \cite{Rozanska+06} for a full discussion. 
Here, we only note that the $T$-discontinuity depends on the local microphysics of the cooling-heating equilibrium, and not on the overall slab structure,  which is the main prediction of RPC.

\subsection{AMD vs. dust content}
\newcommand{\dtom}{\nicefrac{d}{m}}

All the calculations above assume that the ionized AGN outflows have no embedded dust grains. What is the effect of grains on the AMD?

\begin{figure} 
\includegraphics{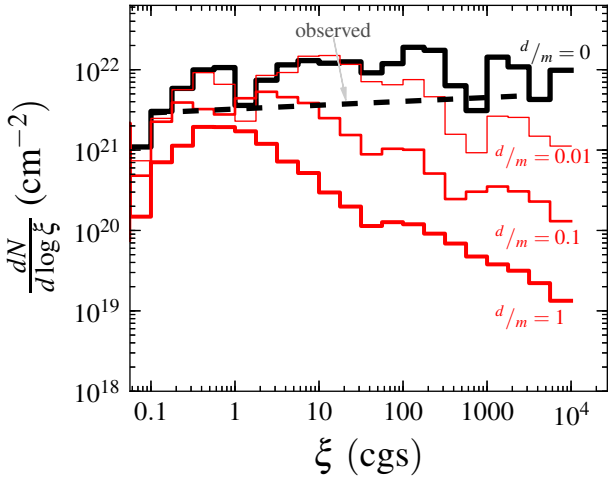}
\caption{
The dependence of the AMD on dust content. 
The solid red lines represent the AMDs of dusty models with different dust to metals ratios (in ISM units). 
The solid black line represents the AMD of the dustless model from Fig. \ref{fig: AMD vs const density}. 
When dust dominates the opacity, the AMD decreases with increasing $\xi$, in contrast with a flat or slightly increasing AMD when gas dominates the opacity. The minimum value of $\xi$ where dust dominates $\sigbar$ increases with decreasing $\dtom$. 
The non-decreasing observed mean AMD (dashed line), implies that ionized AGN outflows are completely devoid of dust, at least in layers with $\xi>3$. 
}
\label{fig: AMD vs dust}
\end{figure}

In Figure \ref{fig: AMD vs dust} we compare the predicted AMDs of slabs with different ratios of dust mass to metal content ($\dtom$), where the ratio is indicated relative to the grain to metal ratio in the ISM. 
We use the ISM grain model available in \cloudy, which includes graphite and silicate components with a size distribution and abundance appropriate for the ISM of the Milky Way. 
In order to conserve the total metals mass, in these dusty \cloudy\ models we reduce the metals depletion on to grains by a factor of $\dtom$\footnote{Some of the small silicates reach temperatures which are above their sublimation temperature at the smaller radii in the simulations. We disregard this complication, which should not affect our conclusions.}.

The spectrum-averaged opacity of the dust grains is $10^{-21}\dtom\cm^2$, which implies that for $\dtom=1$, the dust opacity exceeds the gas opacity at an ionization parameter of $U>0.01$ (\citealt{NetzerLaor93}), equivalent to $\xi>0.3$ for our assumed $\aion=-1.6$. 
Conversely, in dustless gas, $\sigbar<10^{-22}\cm^2$ at $\xi>1$ and $\sigbar<10^{-23}\cm^2$ at $\xi>30$ (app. \ref{app: coeffs}), implying that for $\dtom=0.1$, dust dominates $\sigbar$ at $\xi>1$, while for $\dtom=0.01$ dust dominates $\sigbar$ at $\xi>30$. 
When the grains dominate $\sigbar$, $\sigbar$ is independent of $\xi$, i.e. $q=0$, compared to $q=0.6$ when the gas dominates $\sigbar$ (\S\ref{sec: analytical}). The difference of 0.6 in $q$ implies that the AMD slope should be lower by $0.6$ (eq. \ref{eq: AMD approx}), and therefore we expect the AMD to drop with $\xi$. This drop is clearly seen in the dusty models shown in Fig. \ref{fig: AMD vs dust}. In contrast, when the gas dominates $\sigbar$, $\sigbar$ decreases with $\xi$ and the AMD is flat.

Figure \ref{fig: AMD vs dust} also shows the mean power-law fit to the observed AMDs from Fig. \ref{fig: AMD}. The observed AMD remains flat up to the highest $\xi$ detected, implying strict upper limits on the dust content within the absorbing gas. Gas layers with $\xi>100$ are completely devoid of dust ($\dtom<0.01$), while at $3<\xi<100$ the observed AMD implies $\dtom<0.1$. Only in layers with $\xi\lesssim1$ the AMD is possibly consistent with an ISM-like dust content, although there too, dust is not required. 

\subsection{Fast outflowing components}

The AMDs shown in Fig. \ref{fig: AMD} are the AMDs of the slow component in each object. Some AGN show additional absorption components with higher $v$, with distinctly different AMDs. 
Examples include the $-1900\kms$ component in MCG-6-30-15 (\citealt{Sako+03, Holczer+10}), the $-2600$ and $-4000\kms$ components in NGC~3516 (\citealt{HolczerBehar12}), the $v=-1900\kms$ and $v=-4500\kms$ components of NGC~4051 (\citealt{Steenbrugge+09}), and also the fastest component of Mrk~509 ($v=-770\kms$, \citealt{Detmers+11}). The low and intermediate ionization states are clearly absent in these high-$v$ components, and the column densities at $\log\xi\approx3.5$ tend to be $\approx10^{23}\cm^{-2}$. These relatively narrow AMDs, with $\d N/\d\log \xi$ much larger than expected from RPC, indicate that these high-$v$ components were not compressed by radiation pressure, despite their high $\xi$ indicating that $\prad\gg\pgas$ (eq. \ref{eq: min xi}). 
If radiation pressure did not compress the gas in these fast components, these components are most likely accelerating, which might be related to their high velocity and short lifetime along the line of sight (\citealt{HolczerBehar12}). 

\section{Implications and Discussion}

\subsection{Comparison with R{\'o}{\.z}a{\'n}ska et al. (2006)}
Using the code \titan, \citeauthor{Rozanska+06} (2006, hereafter R06) calculated the slab structure of an ionized AGN outflow which is in hydrostatic equilibrium with the radiation pressure (eq. \ref{eq: hydrostatic only radiation}), hence their models are RPC. 
R06 showed that RPC produces an absorber which spans a wide range of ionization states, as observed in Seyfert outflows. Later, they compared the predicted column densities of ions with observations of NGC~3783 (\citealt{Goncalves+06,Goosmann+11}, Goosmann et al., in prep.), achieving good agreement, thus supporting the RPC picture. 
Our approach is complementary to their approach.  
Rather than fitting the ionic column densities of individual objects, we study the AMD derived from RPC in general. 
We show that the normalization and slope of the AMD is a very robust prediction of RPC, as can be seen from its weak dependence on model parameters (Fig. \ref{fig: AMD vs params}), and as can be understood from simple analytic arguments (\S\ref{sec: analytical}). 
Therefore, the fact that RPC generally reproduces the properties of the observed AMDs (Fig. \ref{fig: AMD}) provides strong evidence that Seyfert outflows are RPC.

\newcommand{\mdot}{{\dot M}}

\subsection{Comparison of RPC with the Krolik et al. (1981) model}

A commonly invoked mechanism to explain the range of $\xi$ observed in Seyfert outflows is the `thermal instability' model, first introduced by \citeauthor{Krolik+81} (1981, hereafter K81). 
In this model, the dense low-$\xi$ gas is confined by high-$\xi$ gas, where the two (or more) `phases' are in pressure equilibrium. The different phases are assumed to be different solutions of the $T$-equilibrium equation, at values of $\Xi\equiv\prad/\pgas$ where the solution is multi-valued. 

\begin{figure}
\includegraphics{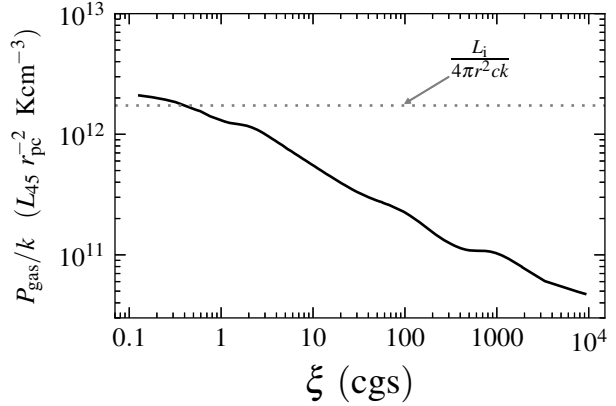}
\caption{
The predicted gas pressure as a function of ionization level in a RPC slab. 
The absorption of radiation momentum causes the gas pressure to increase, and $\xi$ to decrease, with increasing depth into the slab.
The absolute value of $\pgas$ scales as $\prad$, i.e. as $L r^{-2}$. At layers deep enough that most of the radiation pressure has been absorbed $\pgas \approx \prad$ (dotted line). 
Therefore, RPC predicts that low-$\xi$ layers will have higher $\pgas$ than high-$\xi$ layers. 
}
\label{fig: Pgasvsxi}
\end{figure}

The RPC and K81 mechanisms are qualitatively different. 
In the RPC solution, the dense low-$\xi$ gas is confined at the illuminated side by the radiation pressure itself, not by the high-$\xi$ gas. 
Therefore, RPC predicts that $\pgas$ increases with decreasing $\xi$, in contrast with the K81 model where $\pgas$ is constant. 
Figure \ref{fig: Pgasvsxi} presents this trend by plotting the $\pgas$ vs. $\xi$ relation for the RPC slab shown in Figs. \ref{fig: RPCbasic}--\ref{fig: AMD vs const density}. 
Such a trend of increasing $\pgas$ with decreasing $\xi$ has been observed in the ionized outflows of Mrk~279 (\citealt{Ebrero+10}) and in HE~0238-1904 (\citealt{Arav+13}). This distinct feature of RPC should be tested in more AGN outflows in which $\pgas$ can be estimated. 

We note that the term `constant total pressure' used by R06 refers to the term RPC used here, while the term `constant pressure' refers to K81.

\subsection{The dust content of AGN outflows}

\cite{Crenshaw+03} noted that the observed optical reddening in AGN outflows is significantly less than the $\ebv = 2-20$ expected from the observed $\Ntot=10^{22}-10^{23}\cm^{-2}$ and an ISM dust-to-gas ratio, suggesting that the absorbing gas is largely dust free. 
Fig. \ref{fig: AMD vs dust} supports this picture, as it shows that in all layers with $\xi>100$, the dust content is $<1\%$ of the ISM value, while in layers with $\xi>3$ the dust content is $<10\%$ of the ISM value. 

What physical mechanism destroys the dust grains at high $\xi$?
In Paper I, we discuss two processes which destroy dust grains that are relevant to ionized gas in AGN, grain sublimation and grain sputtering (see also \citealt{LaorDraine93,Reynolds97}). 
Grain sublimation strongly depends on $r$, which is not well constrained. 
Grain sputtering is expected to be efficient in gas with $T>10^5-10^6\K$ (\citealt{Draine11b}). 
In RPC, the gas has $T>10^5\K$ at $\xi>100$ (Fig. \ref{fig: RPCbasic}). From eq. 25.14 in \cite{Draine11b} we get that the sputtering timescale is
\begin{equation}\label{eq: sputtering}
 t_{\rm sputtering} \approx \xi T_5^{-3} a_{0.1} L^{-1}_{45} r_{\rm pc}^2 ~~\yr ~~~,
\end{equation}
where $a_{0.1}$ is the grain size in units of $0.1\mic$, $T=10^5T_5\K$, and the dependence of $t_{\rm sputtering}$ on $\nH$ is replaced with a dependence on $\xi$ using eq. \ref{eq: xi}. The relatively short timescales suggested by eq. \ref{eq: sputtering} suggests that the lack of dust grains in $\xi>100$ layers is because the grains are sputtered by the hot gas. 

In gas layers with $\xi<100$, grain sputtering may be efficient if the grains are drifting supersonically relative to the gas. Such drift is possible in layers with $T>10^{4.5}\K$ (section 4.1.2 in Paper I), which corresponds to gas layers with $\xi > 5$. 

However, some AGN with X-ray absorbers show signs of dust along the line of sight in their optical and X-ray spectra. 
\cite{Brandt+96} note $\ebv>0.3$ in IRAS~13349+2438, while \cite{Reynolds+97} note $\ebv=0.61-1.09$ in MCG-6-30-15. Also, \cite{Lee+01,Lee+13} find dust absorption features in the X-ray spectra of these objects, though this result is disputed by \cite{Sako+03}. 
This dust must be associated with gas at {\it some} $\xi$. As Fig. \ref{fig: AMD vs dust} and theoretical considerations exclude the existence of dust in layers of AGN outflows that have high $\xi$, dust in the outflows is likely associated with gas with $\xi\approx0.1-1$, as suggested by \cite{Kraemer+00} and \cite{Ballantyne+03}. 
Another possibility is that dusty gas with $N\approx10^{21}\cm^{-2}$ resides in a distinct absorbing component on the host galaxy scale, where $r$ is large enough so that $t_{\rm sputtering}$ is too long for efficient grain destruction.

\subsection{Open questions}

\subsubsection{Confinement at the shielded side}\label{sec: acceleration}
\newcommand{\fac}{f_{\rm ac}}

The RPC solution is valid for a non-accelerating outflow. 
What then counteracts the outward force of the radiation $\prad$?
If the outflow is plowing into a stationary, low-density medium, then a plausible mechanism which absorbs the radiation momentum is ram pressure on the leading edge of the slab
\begin{equation}
 P_{\rm ram} = \mu \mp n_{\rm amb} v^2 ~~~,
\end{equation}
where $v$ is the outflow velocity, $n_{\rm amb}$ is the density of the ambient gas, and $\mu\mp$ is the gas mass per H-nucleon. 
For $P_{\rm ram}$ to balance the gas pressure on the shielded side, which is $\prad\tautot$ (eq. \ref{eq: pgas leading edge}), $n_{\rm amb}$ needs to be:
\begin{equation}
 n_{\rm amb} \approx  7 \times 10^4 ~\mu^{-1}\tautot L_{45} r_{\rm pc}^{-2} v_{500}^{-2} \cm^{-3} ~~~,
\end{equation}
where $v=500\,v_{500}\kms$. 
Such gas will have $\xi=1600\,\mu \tautot^{-1} \left(\frac{v}{500\kms}\right)^2$, and therefore in principle could be observable. 
However, hydrodynamical simulations are required to constrain what fraction of the total volume is filled by this confining gas, and whether its column density is large enough for it to be observable. Note that if $\tautot\lesssim0.1$, then the ambient medium is fully ionized, and will not be observable.

\subsubsection{Hydrodynamical stability of the RPC solution}

A remaining open question is the stability of a radiation compressed outflow.
This subject was addressed in studies for various configurations related
to AGN (\citealt{MathewsBlumenthal77, BlumenthalMathews79, Mathews83, Mathews86, Krolik79, Krolik88, Jiang+13, Davis+14}).
A possibly significant instability source is Rayleigh-Taylor instability,
which occurs when the density gradient and the acceleration are in opposite
directions. If the outflow is accelerated outwards by the radiation
pressure, and part of this radiation is balanced by the built up gas pressure (i.e. the
hydrostatic RPC solution), then the lower density gas lies ``above'' the
higher density outflowing gas, and the flow should be  Rayleigh-Taylor stable 
(see, e.g., the analysis by \citealt{Kuiper+12} on the interaction of radiation pressure with accreting gas in massive stars). 
However, as mentioned above, the outflowing gas is likely plowing into a low density ambient
medium, and thus the leading face of this outflow forms a dense medium
above a low density medium, which is expected to be Rayleigh-Taylor unstable.
The timescale for the development of this instability, and its effect on the
outflow require numerical simulations which can follow the growth of the
instability beyond the linear growth phase (\citealt{Jiang+13,Davis+14}). 

This Rayleigh-Taylor instability may be relevant for understanding the
velocity dispersion in the outflows. The RPC solutions correspond to gas
outflowing at a constant velocity, which will produce only thermal broadening of the absorbing gas,
while typically AGN outflows feature super-thermal broadening (see also app. \ref{app: pline}). The
instability may induce a supersonic large scale turbulence at the leading edge
of the outflow, which would broaden the lines beyond the thermal width. 
Numerical simulations can be used to test this possible mechanism.

The RPC outflow is convectively stable, as hot low-pressure gas overlies
dense high-pressure gas, while convective instability requires a negative slope of
the entropy with pressure. 

The outflowing gas is also  Kelvin-Helmholtz unstable to the shear against the
ambient gas in its lateral boundaries. This instability is relevant to any
mechanism which drives outflowing filaments of gas. Again, a quantitative estimate of
the impact of this instability requires detailed numerical simulations.

\subsubsection{The mass outflow rate}
One long-sought property of AGN outflows is their mass outflow rate $\mdot$, which can determine the extent of feedback exerted by the AGN on its host galaxy, a mechanism often employed in structure formation models (e.g. \citealt{Hopkins+06}). In order to evaluate $\mdot$, one needs an estimate of $r$ (e.g. \citealt{Crenshaw+03,CrenshawKraemer12,Arav+13}). However, the insensitivity of the AMD to $r$ (Fig. \ref{fig: AMD vs params}) indicates that the AMD is not directly useful to constrain $r$, and therefore the typical values of $\mdot$ remain an open question.

\subsection{The advantage of the AMD method over multi-component iso-$\xi$ fits}

Standard fitting models, e.g., \xstar\ models in \xspec, for the sake of simplicity use iso-$\xi$ components -- the opposite of the broad distribution implied by RPC. Since for practical purposes any continuous distribution can be approximated by a discrete set of components, it is advisable to use these models by invoking a series of well defined ionization components that span the ionization range of $-1 < \log \xi < 4$ expected from RPC. Since each ion forms over a range of $\xi$, there is an empirical limit to the AMD resolution (in $\xi$) that one can expect to achieve from fitting spectra. Usually, $4-6$ ionization components are sufficient to cover the observed range of ionization states, and using many more would only result in a degeneracy between adjacent components. 
Even when fewer than four components achieve a statistically reasonable fit to the spectra in terms, say, of reduced $\chi^2$, the present work shows that a continuous distribution provides a more physically meaningful picture of the outflow structure. Additionally, 
by using several components, observers can recover the full AMD including its possible gaps (instabilities), which the statistically based trial-and-error fitting approach of adding iso-$\xi$ components is obviously not able to recover. The number of free parameters can be reduced by using an analytic power-law fit for the AMD, which has four free parameters (minimum $\xi$, maximum $\xi$, slope, and normalization), compared to $8-12$ free parameters needed to fit $4-6$ individual clouds ($\xi$ and $N$ for each).

We re-iterate from \cite{Holczer+07} that the AMD insight into the outflow structure was inspired by a similar approach commonly used in the analysis of astrophysical plasma (emission) sources starting from the Sun (\citealt{ShmelevaSyrovatskii73}), through stellar coronae (\citealt{GudelNaze09}), and all the way to cooling flow models of galaxy clusters (\citealt{PetersonFabian06}). Ionization distributions, are thus a tool to understand the physics of the plasma at hand more than they are an empirically fitting method.

\section{Conclusions}
The ionization levels of AGN outflows imply that $\prad\gg\pgas$. 
Since the transfer of energy from the radiation to the gas is always associated with a transfer of momentum, 
the radiation will either compress or accelerate the absorbing gas. 
We solve the hydrostatic slab structure of radiation pressure compressed (RPC) gas both analytically and numerically. 
We show that a single slab of RPC gas produces a typical density profile of $n \propto x^{-1}$ inside the slab, and spans a wide range of ionization states, as found previously by \cite{Rozanska+06}. 
Additionally, we find that: 
\begin{enumerate}
 \item RPC predicts a robust normalization and slope of the Absorption Measure Distribution: $\d N/\d\log\xi \approx 7.6\times10^{21}\xi^{0.03}\cm^{-2}$. 
 \item The normalization and slope of the AMD are independent of $\Ntot$, $\xi_0$, and $r$, and only weakly depend on $Z$ and $\aion$. 
 \item The theoretical slope agrees with observations of AGN outflows with $v$ of a few $100\kms$, while the normalization is smaller by a factor of three, possibly due to atomic data uncertainties.
 Therefore, the observed AMDs of AGN outflows strongly suggest that the outflows are RPC. 
 \item The AMD slope in RPC gas depends on its dust content. The observed slopes indicate that AGN outflows are largely devoid of dust. 
 \item RPC has a unique signature that $\pgas$ decreases with increasing $\xi$, in contrast with a multiphase absorber in pressure equilibrium. This prediction can be tested in future observations. 
\end{enumerate}

\section*{Acknowledgements}
We thank G. Ferland and his team for developing \cloudy, making it publicly available, and providing excellent support. This work also makes use of the \xstar\ code, kindly made available by T. Kallman. We thank Steven B. Kraemer, Ren{\' e} Goosmann, Fabrizio Arrigoni Battaia, and the referee, Agata R{\'o}{\.z}a{\'n}ska, for constructive and insightful comments. 
E.B. is supported by the I-CORE program of the Planning and Budgeting Committee and the Israel Science Foundation (grant numbers 1937/12 and 1163/10), and by a grant from Israel's Ministry of Science and Technology.
A.L. acknowledges support through grant 2017620 from the Helen and Robert Asher Fund at the Technion. 

\appendix

\section{$T(\xi)$ and $\sigbar(\xi)$}\label{app: coeffs}

\begin{figure} 
\includegraphics{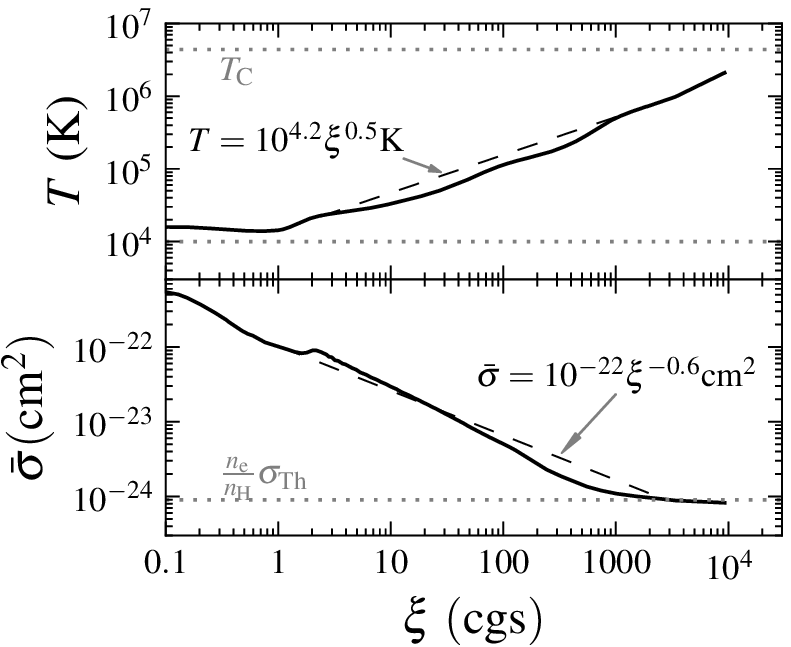}
\caption{
A comparison of $T$ and $\sigbar$ with $\xi$, for each zone in the \cloudy\ calculation of the model shown in Figs. \ref{fig: RPCbasic} -- \ref{fig: AMD vs const density}.
The value of $T$ drops from near $\tc$ at $\xi=10^4$, to $T\approx10^4\K$ at $\xi=1$. 
The value of $\sigbar$ increases from the electron scattering opacity at $\xi=10^4$ to $\sigbar=10^{-22}\cm^{2}$ at $\xi=1$. 
Dashed lines show the power-law interpolations of $T(\xi)$ and $\sigbar(\xi)$ between $\xi=1$ and $\xi=3000$, which are used in the semi-analytic approximation of the AMD (eq. \ref{eq: AMD estimate} and Fig. \ref{fig: AMD vs const density}). 
}
\label{fig: coeffs}
\end{figure}

Figure \ref{fig: coeffs} compares $T$ and $\sigbar$ with $\xi$ for the RPC model shown in Figs. \ref{fig: RPCbasic} -- \ref{fig: AMD vs const density}. The values of $T$, $\sigbar$ and $\xi$ are calculated by \cloudy\ in each zone of the slab. 
The value of $T$ drops from $2\times10^6\K$ at $\xi=10^4$, to $T=10^{4.2}\K$ at $\xi=1$. The drop is due to the lower $\prad/\pgas$ at lower $\xi$, which implies a heating-cooling equilibrium at lower $T$. 

The lower panel shows that $\sigbar$ increases from the electron scattering opacity $(n_{\rm e}/\nH)\sth=10^{-24.1}\cm^2$ at $\xi=10^4$, where $\sth$ is the Thomson cross section, to $\sigbar=10^{-22}\cm^{2}$ at $\xi=1$. 
At $1<\xi<3000$, both $T(\xi)$ and $\sigbar(\xi)$ can roughly be approximated as power-laws. We plot the interpolations between the values of $\sigbar$ and $T$ at $\xi=1$ and $\xi=3000$. These power-law interpolations are used in the semi-analytic approximation of the AMD (eq. \ref{eq: AMD estimate} and Fig. \ref{fig: AMD vs const density}).

\section{Non-thermal pressure terms}\label{app: pline}

The hydrostatic equilibrium equation (\ref{eq: hydro simple form}) used above to derive the AMD assumes that the magnetic pressure $\pmag$ and the pressure of the trapped line emission $\pline$ are negligible compared to $\pgas$. 
If we include these pressure terms, eq. \ref{eq: hydro simple form} takes the form
\begin{equation}\label{eq: with pline}
 \d (\pgas + \pline(\vturb) + \pmag) = \prad e^{-\taubar}\d\taubar ~~~,
\end{equation}
where we explicitly noted the dependence of $\pline$ on the velocity dispersion within the absorber $\vturb$.

We first address the effect of a finite $\pline$. Observations show that X-ray absorption features have velocity widths of hundreds of $\kms$. This highly supra-thermal velocity is unlikely to be turbulent and add to the total pressure, as this would correspond to supersonic motion which would create shocks that heat the gas to $>10^6\K$. More likely, the velocity widths are dominated by large scale ordered motion in the absorbing gas. Therefore, the exact relation between the observed widths and the $\vturb$ which enters eq. \ref{eq: with pline} is not entirely clear. 

\begin{figure} 
\includegraphics{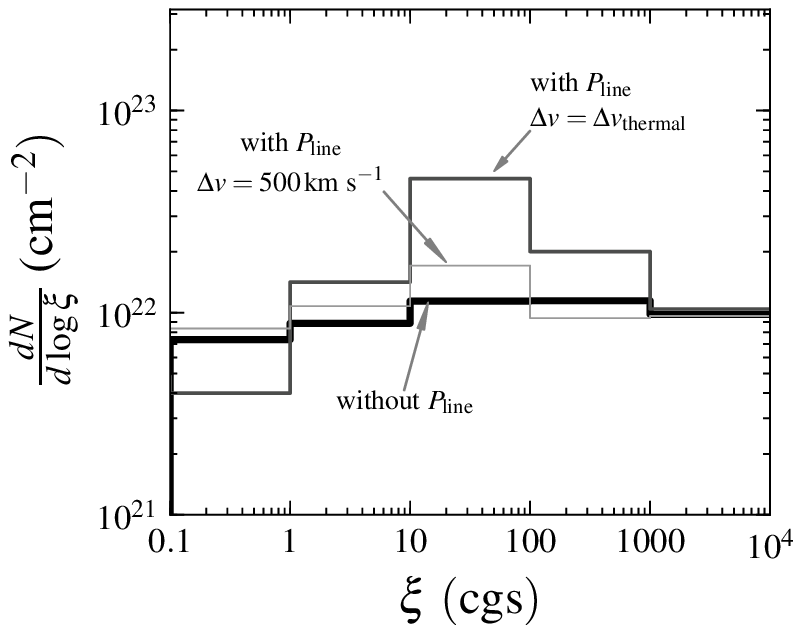}
\caption{The effect of $P_{\rm line}$ on the AMD. 
Three AMDs calculated by \cloudy\ are shown, with $\pline$ turned off (thick line), 
with $\pline$ turned on and assuming thermal line broadening (intermediate-width line), 
and with $\pline$ turned on and assuming $\vturb=500\kms$, the typical observed velocity width of X-ray absorption features (thin line). 
In the model with $\pline$ and $\vturb=\Delta v_{\rm thermal}$, the AMD at $10<\xi<1000$ is higher by a factor of $2-4$ compared to the model without $\pline$. 
The AMD of the model with $\pline$ and $\vturb=500\kms$ is similar to the AMD of the model without $\pline$. 
}
\label{fig: line pressure}
\end{figure}

Figure \ref{fig: line pressure} shows the effect of $\pline$ on the AMD (see also appendix B in Paper II). For clarity, we plot the AMDs in decade-wide bins in $\xi$. 
Assuming only thermal broadening, including $\pline$ causes the AMD at $10<\xi<1000$ to be higher by a factor of $2-4$. 
This increase in the AMD occurs since the absorbed $\prad$ is counteracted partially by the increase in $\pline$ and partially by the increase in $\pgas$, compared to entirely by $\pgas$ in the previous RPC model where $\pline$ is turned off. 
The implied more gradual increase in $\pgas$ (and in $\nH$) with $N$, implies a more gradual decrease in $\xi$, which explains the larger AMD in the model which includes $\pline$. 

However, Fig. \ref{fig: line pressure} also shows that when assuming $\vturb=500\kms$, the lines are not trapped and $\pline$ is low enough so the AMD is similar to the model without $\pline$. 
Therefore, when assuming that $\Delta v$ equals the observed widths of X-ray absorption features, our choice of neglecting $\pline$ is justified. 
However, due to the currently unknown physical source of the velocity field in AGN outflows, the effect of $\pline$ on the AMD remains an open question. 

A $\pmag$ which is $\gg\pgas$ will also increase the expected value of the AMD.
Indeed, compression of gas can enhance the magnetic field, e.g. when a star collapses to a neutron star. 
However, in RPC Seyfert outflows, both the initial strength of the magnetic field and the initial gas density before compression are not constrained, so $\pmag$ is a free parameter. 
The correspondence between the observed AMD and the predicted AMD assuming $\pmag=0$ (Fig. \ref{fig: AMD}), suggests that in Seyfert outflows $\pmag$ is not much larger than $\pgas$.  
We note in passing that a study of the effect of $\pmag$ on ionized gas in star forming regions was conducted by \cite{Pellegrini+07}. 

\section{Comparison with \xstar}\label{app: xstar}

In order to derive an estimate of how the calculated AMD depends on the systematics of the radiative transfer calculations, we compare an AMD calculation by \cloudy\ with an AMD calculation based on \xstar\ (\citealt{KallmanBautista01}). 
We use the analytical {\it warmabs} model\footnote{http://heasarc.gsfc.nasa.gov/xstar/docs/html/node99.html} version 2.1ln8 of \xstar, which is available through \xspec\footnote{http://heasarc.gsfc.nasa.gov/docs/xanadu/xspec/models/xstar.html}.

The {\it warmabs} calculation is limited to $\aion=-1$, and a single ionization state and $\pgas$ for each $\xi_0$ value, so no radiation transfer is computed in each individual run. Therefore, in order to obtain the AMD, we calculate $\sigbar(\xi)$ from the {\it warmabs} calculations of the transmission spectra of a discrete set of $\log \xi$ values, over the spectral range of $1 - 1000\ryd$. We use the values of $T(\xi)$ derived by \cite{Holczer+07}. 
The AMD is then derived by chaining together {\it warmabs} results for the discrete set of $\xi$ values, according to eq. \ref{eq: AMD}. We use $Z=\zsun$ and \cite{GrevesseSauval98} abundances.

\begin{figure} 
\includegraphics{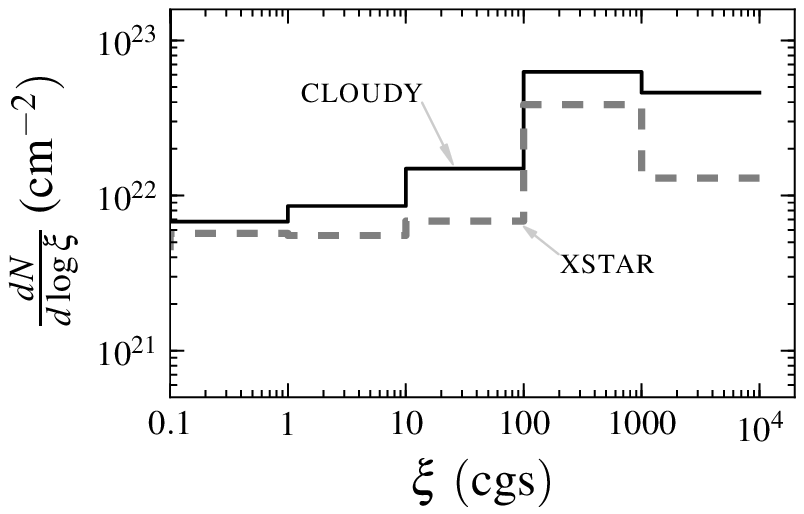}
\caption{
A comparison between the AMD calculations of \cloudy\ and \xstar. 
The \xstar\ and \cloudy\ AMD calculations agree to within a factor of three. 
}
\label{fig: xstar}
\end{figure}

Fig. \ref{fig: xstar} compares the AMD calculated by \xstar\ with an AMD calculation by \cloudy\ for the same $\aion,\ Z$, abundances, and granularity in $\log \xi$ as the {\it warmabs} calculation. Due to an energy conservation problem in \cloudy\ when using $\aion=-1$, the spectral slope at $h\nu>10\kev$ was set to $-2$. Fig. \ref{fig: xstar} shows that the two calculated AMDs agree to within a factor of three.
The differences are likely due to the inclusions of different absorption lines and to the different ionization balance. 
For example, if \xstar\ has more metal lines in the high-$\xi$ regime, then at these $\xi$, $\sigbar$ will be higher in the \xstar\ calculation than in the \cloudy\ calculation, which can explain the lower AMD (eq. \ref{eq: AMD}).

\end{document}